\documentclass{aastex}

\shorttitle{New SX Phoenicis Stars } \shortauthors{Jeon et al.}

\begin{document}

\title{New SX Phoenicis  Stars in the Globular Cluster M53 \\
}

\author{Young-Beom Jeon\altaffilmark{1}}
\affil{Korea Astronomy Observatory, Daejeon 305-348, KOREA \\
Email: ybjeon@boao.re.kr}

\author{Myung Gyoon Lee}
\affil{Astronomy Program, School of Earth and Environmental Sciences,
Seoul National University, Seoul 151-742, KOREA \altaffilmark{2} and \\
Department of Terrestrial Magnetism, Carnegie Institution of
Washington, 5241 Broad Branch Road, N.W., Washington, D. C. 
\\ Email: mglee@astrog.snu.ac.kr}

\author{Seung-Lee Kim}
\affil{Korea Astronomy Observatory, Daejeon 305-348, KOREA \\
Email: slkim@kao.re.kr}

\and

\author{ Ho Lee}
\affil{Department of Earth Science Education, Korea National University of
Education, Choongbuk 363-791, Korea\\
Email: leeho119@boao.re.kr}

\altaffiltext{1}{Also Astronomy Program, School of Earth and Environmental Sciences,
Seoul National University, Seoul 151-742, KOREA }
\altaffiltext{2}{Permanent address}

\begin{abstract}

Through time-series CCD photometry of the metal-poor globular cluster M53,
we have discovered eight  new SX Phoenicis type stars (labeled from SXP1 to SXP8).
All the new SX Phoenicis stars are located in the blue straggler star
region of a color-magnitude diagram of M53.
One of these stars (SXP2) is found to have very closely separated pulsation frequencies:
$f_1/f_2 = 0.9595$
where $f_1$ and $f_2$ are primary and secondary frequencies.
This may be due to excitation of non-radial modes.
Six of these  SX Phoenicis stars are considered
to be pulsating in the fundamental mode.
They show a tight linear correlation between the period and luminosity.
We derive a period - luminosity relation for the fundamental mode
for the period range of $-1.36 < Log P[d]< -1.15$ :
$<V>=-3.010(\pm0.262)Log P + 15.310(\pm0.048 )$ with an rms scatter of 0.038,
corresponding to $<M_V>=-3.010 Log P - 1.070$ for an adopted distance modulus
of $(m-M)_V=16.38$ (Harris 1996).
\end{abstract}


\section{Introduction}

SX Phoenicis stars are Population II pulsating variable stars with
shorter periods ($<0^d.1$) and  larger amplitudes of variability,
compared with $\delta$ Scuti stars which are short period pulsating variable stars
belonging to Population I.
Interestingly most of known SX Phoenicis stars in globular clusters are
located in the blue straggler star (BSS) region
in the color-magnitude diagram (CMD).
It is long since the presence of the BSS in the globular cluster M3 became known
by \citet{san53}. Now there are many BSS discovered in the globular clusters
and open clusters in our Galaxy and in the fields of nearby galaxies.
However, the origin of the BSS is still controversial.
The BSSs are hotter and brighter than the main sequence turn-off stars so that
it is needed some mechanisms to make the stars hotter and brighter
(more massive) to explain the origin of the BSS.
There have been suggested two classes of mechanisms to explain the origin of the BBS.
One class involves a single star where mixing in the atmosphere can increase the life time
of the main sequence. The other class is based on merger process where two low-mass stars
are merged to form a more massive unevolved star via mass transfer or direct collisions
between stars. These days the latter is preferred \citep{bai95}. 
Recently a significant fraction of the BSS in globular clusters are known to be
SX Phoenicis stars. The presence of  these pulsating variable stars (as well as eclipsing variables)
among the BSS provides an excellent opportunity to understand the origin of the BSS,
because we can investigate physical properties and processes  of the stars in more detail compared
with the case of non-variable stars.
Therefore the study of SX Phoenicis stars can
 provide us with important clues on the origin of the BSS.

It is only after 1980's that SX Phoenicis stars were discovered in the globular clusters.
Since the first discovery of SX Phoenicis stars in the globular cluster
$\omega$ Centauri \citep{nis81},
the discovery rate of these stars
in the globular clusters has increased rapidly in the last decade, using the
intermediate-to-large telescopes.
Recently, many SX Phoenicis stars are being discovered
in several globular clusters and nearby galaxies.
Some examples are
$\omega$ Centauri \citep{kal97}, M5 \citep{kal99}, M15 \citep{jeo21},
M22 \citep{kal01}, 47 Tuc \citep{gil98,bru01}, and M55 \citep{pyc01}.
\citet{rod20} listed a total of 122 SX Phoenicis stars
belonging to 18 globular clusters and 27 stars
in two external galaxies, covering  information published until January, 2000.
However, 
the characteristics of the SX Phoenicis stars
have not yet been fully explained by the present stellar
evolution theories, requiring further investigation in observation and theory.

In addition, SX Phoenicis stars are known to show period-luminosity (P-L)
or period-metallicity-luminosity (P-[Fe/H]-L)
relation, which can be a very useful distance indicator for globular clusters
and nearby galaxies \citep{mcn97,nem94}.
However, the number of known SX Phoenicis star in a
given globular cluster is small  and most studies use an inhomogeneous sample
based on different sources.  Therefore there is seen a large variation among
these relations.  For example, \citet{mcn97} derived a slope of --3.7, while
\citet{pyc01} obtained a much flatter slope of --2.9.

In 1999 we began time series CCD photometry to search for
variable BSSs in globular clusters
using the 1.8m telescope at the  Bohyunsan Optical Astronomy Observatory (BOAO)
in Korea.
Generally we can detect SX Phoenicis type stars brighter than $V=19.5$ mag for
200 seconds exposures, depending on the seeing condition.
For larger amplitude variable stars such as W UMa type eclipsing binaries,
the detecting magnitude reaches 20.2 mag for the same exposures.
The first result of our survey was discoveries of an SX Phoenicis star \citep{jeo21}
and two W UMa type variable stars \citep{jeo22} in the globular cluster M15.

In this paper we present the result of our survey for short period variable stars in the
metal-poor globular cluster M53.
M53 (RA\,=\,13$^h$ 12$^m$ 55$\fs$3, DEC\,=\,+18$\arcdeg$ 10$\arcmin$
09$\arcsec$, J2000.0; \citet{har96}) has a very  low
metallicity [Fe/H] $=-1.99$, a low interstellar reddening
$E(B-V)=0.02$, and an apparent $V$ distance modulus $(m-M)_V=16.38$
\citep{har96}.
Many blue straggler stars are known to exist in M53 \citep{rey98}.
Some of these blue stragglers are expected to be SX Phoenicis stars,
but no SX Phoenicis stars have been reported yet \citep{rod20,cle21}.
Here we report a discovery of eight SX Phoenicis stars
in this cluster.

This paper is composed as follows. Observations and data reduction are described in Section 2.
Section 3 presents the light curves of the new SX Phoenicis stars in M53 and
Section 4 discusses the characteristics of these stars, including
the pulsation modes and the period-luminosity relation.
Finally primary results are summarized in Section 5.

\section{Observations and Data Reduction}

\subsection{Observations}

We obtained time series CCD images of M53 for eleven nights from
March 12th, 1999 to April 1st, 2001.
A total of 45 and 381
frames were obtained for $B$ and $V$ bands,
respectively.
Because the observations were performed under various seeing condition,
we adjusted exposure times depending on the seeing.
 The observation log is listed in Table 1.
\placetable{tbl-1}

The CCD images were obtained with a thinned SITe 2k CCD
camera attached to the BOAO 1.8m telescope.
The field of view of a CCD image is $11\farcm6\times11\farcm6$
 ($0\farcs3438$ pixel$^{-1}$)
at the f/8
Cassegrain focus of the telescope. The readout noise, gain and readout time of
the CCD are 7.0 $e^-$, 1.8e $^-/$ADU and 100 seconds, respectively.

A greyscale map of a $V$-band CCD image is shown in Figure 1.
It shows only
south-east region ($7\farcm6\times6\farcm7$) of the cluster, out of the total observed
field of $11\farcm6\times11\farcm6$.
Eight new SX Phoenicis stars are represented by circles and labeled from SXP1 to SXP8.
\placefigure{fig1}
\subsection{Data Reduction}

Using the IRAF/CCDRED package, we processed the CCD images to correct
overscan regions, trim unreliable subsections, subtract bias
frames and correct flat field images. Instrumental magnitudes
were obtained using the point spread function (PSF) fitting photometry
routine in the IRAF/DAOPHOT package \citep{mas92}.
The instrumental magnitudes of the stars in M53
were transformed to the standard system using
the $BVI$ standard stars \citep{lan92} observed on the photometric night of March 30, 2000.
In Figure 2 we plotted CMDs for a total of
about 18,000 stars on the ($V$,$B-V$) plane.
The left panel in Figure 2 shows the CMD for a central region at r$ < $1$\farcm0$,
and the right panel shows the CMD for an outer region at r$ \ge$1$\farcm$0.
\placefigure{fig2}
On the right panel, the main sequence (MS), the red giant branch (RGB)
 and the horizontal branch (HB) are clearly seen. In addition, there are
 about 100 stars at the brighter and bluer region above the MS turnoff
 (represented by a box), which are blue stragglers.
Large scatters shown in the left panel CMD for r$<$1$\farcm0$ are due to
crowding, so that we could not find any variable stars at  $r<$1$\farcm0$.

We applied the ensemble normalization technique
\citep{gil88,jeo21} to normalize instrumental magnitudes between
time-series CCD frames. We used about a hundred  normalizing
stars ranging from 13\fm7 to 17\fm8 for the $V$-band and from 13\fm7
to 18\fm5 for the $B$-band, except for variable stars and central
stars at r $<$ 1\farcm0 where the crowding is severe.
For   $B$-band data, we use them only for obtaining mean magnitudes, because
the data quality was not good enough to apply frequency analysis.
The normalization equation is
\begin{equation}
  B ~ or ~ V = m + c_1 + c_2(B-V) + c_3P_x + c_4P_y
\end{equation}
where $B$, $V$, and $m$ are the standard  and instrumental
magnitudes of the normalizing stars, respectively. $c_1$ is the zero
point and $c_2$ is the color coefficient. $c_3$ and $c_4$ are
used to correct position-dependent terms such as  atmospheric
differential extinction and  variable PSF.

\section{Light Curves and Frequency Analysis}

After photometric reduction  of the time-series frames,
we inspected luminosity variations for about 18,000 stars
to search for variable stars.
>From the variable star search
we discovered eight new SX Phoenicis stars and recovered about tenth of  previously
known RR Lyrae stars.
Here we report only the results on the SX Phoenicis stars.
Detailed results on the RR Lyrae stars
will be presented in a separate paper \citep{jeo03}.

The mean magnitudes and color indices of the SX Phoenicis stars are listed in Table 2.
The R.A.(J2000.0) and Decl.(J2000.0) coordinates of the stars in Table 2
were obtained from the astrometry using the Guide Star Catalogue (Version 1.1).
Interestingly all the new SX Phoenicis stars are located in the south-east direction
from the center of M53.
We checked any possible systematic errors preventing us from finding
variable stars in other region of the field, but finding none.
The coordinates, mean magnitudes and color indices of the eight
new SX Phoenicis stars are listed in Table 2.
\placetable{tbl-2}

Figure 3 displays
$V$-band light curves of eight new SX Phoenicis stars.
Because SXP3 is located near the cluster center and SXP4 close to the CCD edge,
some data of SXP3 were lost by poor seeing and some data of SXP4 by daily variation
of observing field center.
The light curves in Figure 3 show  typical characteristics of SX Phoenicis stars, i.e.,
short periods and low amplitudes.
\placefigure{fig3}

We have performed multiple-frequency analyses to derive the pulsating
frequencies of  the eight SX Phoenicis stars, using
the discrete Fourier transform
and  linear least-square fitting methods
\citep{kim96}. Figure 4 displays the power spectra 
 the eight SX Phoenicis stars we derived.
Small inner panels in Figure 4 represent the pre-whitening processes.
We searched for one to three pulsating frequencies for each star at the frequency region
from 5 to 70 cycles day$^{-1}$.
Low frequencies detected for the SXP1, SXP3, SXP4 and SXP8 might have resulted
from variable seeing condition and/or drift during long observing runs from 1999 to 2001.
Synthetic light curves obtained from these analyses are
superimposed on the data in Figure 3, fitting the data well.
\placefigure{fig4}

The results of the multiple-frequency analyses for  the eight SX Phoenicis stars
are summarized  in Table 3.
In Table 3, we accepted as intrinsic pulsating frequencies with
their amplitude signal-to-noise ratios larger than 4 \citep{bre93}
except for a frequency $f_3$ of SXP2.
Although the amplitude signal-to-noise ratio of SXP2, 3.8, is slightly smaller than
4, its $f_3$ seems to be a harmonic frequency
of $f_1$.
The periods of primary modes for these SX Phoenicis stars range from 0.0385 days to 0.0701 days,
and the semi-amplitudes of the variability range from 0.030 mag to 0.118 mag.
\placetable{tbl-3}

\subsection{The Characteristics of the individual SX Phoenicis star}

{\bf SXP1:} SXP1 has the longest primary period of 0.0701 days
   among the eight new SX Phoenicis type stars of M53.
   This star has two frequencies, $f_1=14.2600$ cycles day$^{-1}$ and $f_2=27.5256$ cycles day$^{-1}$.
   This star might be a mono-periodic pulsator,
   because the second frequency $f_2$ seems to be caused by 1 cycle day$^{-1}$
   alias effect of a harmonic frequency of $f_1$, $2f_1=28.5200$ cycles day$^{-1}$,
   In Figure 3, the light curves of SXP1 shows an asymmetric sinusoidal feature,
   which is the characteristic of harmonic frequencies.
   The maximum amplitude of SXP1 is estimated to be $\Delta V=$ 0.292 mag.

{\bf SXP2:} Light curves of this star in Figure 3
   show an amplitudes change from day to day,
   implying the excitation of multiple pulsating frequencies.
   We derived two closely separated frequencies of $f_1=22.0450$ cycles day$^{-1}$
   and $f_2=22.9750$ cycles day$^{-1}$, with a ratio of $f_1/f_2 = 0.9595$.
   These frequencies can be explained by excitation of  non-radial modes.
   Pulsations with non-radial modes were found in several recent observational results
   of the SX Phoenicis stars:
   SX Phoenicis itself \citep{gar96}, BL Cam \citep{zho99}, V3 in 47 Tuc \citep{gil98},
   SXP1 in M15 \citep{jeo21}, and ten SX Phoenicis stars in M55 \citep{pyc01}.
   After these two frequencies are pre-whitened,
   the third frequency
   is detected at $f_3=43.0190$ cycles day$^{-1}$, which might have resulted from 1 cycle day$^{-1}$
   alias effect of $2f_1=44.0900$ cycles day$^{-1}$.
   Even if the amplitude signal-to-noise ratio  for the third frequency is only 3.8,
   the light curves of SXP2 show asymmetric sinusoidal feature, supporting
   the existence of harmonic frequencies.
   The maximum amplitude of SXP2 is estimated to be
   $\Delta V=$ 0.372 mag, which is the largest value
   among the eight SX Phoenicis stars in M53.

{\bf SXP3, SXP4, SXP5, SXP6, SXP7:}
   We could detect only a primary frequency for these five stars.
   Primary frequencies and the maximum amplitudes
   of SXP3, SXP4, SXP5, SXP6 and SXP7 are $f_1=$20.8796, 20.0114,
   22.9880, 22.6000 and 24.1024 cycles day$^{-1}$
   and $\Delta V=$ 0.204, 0.138, 0.082, 0.200 and 0.194 mag, respectively.
   In Figure 4 SXP4 and SXP5 show some hints for the existence of harmonic frequencies
   above 40 cycles day$^{-1}$, but they are not conclusive.

{\bf SXP8:}
   SXP8 has the shortest primary period of 0.0385 days and
   the smallest semi-amplitude of 0.030 mag
   among the eight SX Phoenicis stars of M53.
   SXP8 shows a typical feature of low amplitude $\delta$ Scuti stars (LADS), i.e.,
   the feature of
   complicated oscillation pattern and several frequencies compared to
   one or two stable  frequencies of high amplitude $\delta$ Scuti stars (HADS).
   We could detect three frequencies with amplitude signal-to-noise ratios over 4.0,
   although the data are not excellent. 
   The primary, secondary and third frequencies are
   $f_1=25.9800$, $f_2=38.1942$ and $f_3=8.2322$ cycles day$^{-1}$, respectively.

\section{Discussion}

\subsection{ Characteristics of the New SX Phoenicis Stars }

In Figure 5, we show the position of the eight new
SX Phoenicis stars in the color-magnitude diagram of M53.
All the SX Phoenicis stars discovered in M53 are found to be located
in the blue straggler region,
brighter and bluer than the main sequence turnoff point.
It is interesting that all of them are located in the red side in the
BSS region of M53. This indicates that the pulsational instability strip
may cover only a part of the BSS region (a hotter region in this case)
and that only those (or some of them) in this region can start
pulsation. It is not yet known why some BSSs pulsate and some do not
in the same pulational instability region. It needs further
studies using several globular clusters with SX Phoenicis stars.

\placefigure{fig5}

In Figure 6 we have compared the $V$-amplitudes and
periods of the SX Phoenicis stars in M53 with those of
SX Phoenicis stars and  $\delta$ Scuti stars in other globular clusters
and field SX Phoenicis stars.
The sources of the data in Figure 6
are \citet{rol20} for field SX Phoenicis stars and  $\delta$ Scuti
stars, and \citet{rod20} for SX Phoenicis stars in  Galactic
globular clusters.
Figure 6 shows that the $V$-amplitudes and
periods of  the eight SX Phoenicis stars (star symbols)
are consistent with those for other SX Phoenicis
stars in  globular clusters.  The $V$-amplitudes of
the SX Phoenicis stars are much larger than those of  $\delta$ Scuti stars with
the same period, and the periods of the SX Phoenicis stars
the shortest end.
\placefigure{fig6}

\subsection{ Mode Identification and Period-Luminosity Relation }

The period-luminosity (P-L) relation of SX Phoenicis stars in the globular
clusters can be very useful for estimating the distances to the clusters and nearby
galaxies.
However, it is not easy to derive the P-L relation from the observations,
because there are often seen a mixture of different pulsation modes
in the SX Phoenicis stars.
Different pulsation modes of the SX Phoenicis stars follow different P-L relation
(offsets in the zero points).

It is well known that observational identification
of the pulsation modes of the SX Phoenicis star is difficult in general.
McNamara (1997, 2000, 2001) suggested that the light amplitude and the degree of asymmetry
of the light curves might be useful parameters in deciding the pulsating modes
of the high amplitude  stars.
McNamara (2000) showed that most of the known SX Phoenicis stars at the fundamental mode show
large amplitude (for example, larger than 0.15 mag in the case of SX Phoenicis stars
in $\omega$ Centauri) and asymmetric light curve.
However, not all of the SX Phoenicis stars with large amplitude are at the fundamental mode,
as shown by McNamara (2000).
On the other hand,\citet{pyc01} pointed out ``Amplitudes generally yield no definitive clues for
the identification of modes, except that large amplitudes are more likely to occur in radial
pulsations''.
Another possible way of identifying the pulsating modes is to use the period-luminosity
relations and the period ratios of the SX Phoenicis stars in globular clusters,
being applied in this study.


In Figure 7 we display the period and mean $V$ magnitude relation for the
SX Phoenicis stars in M53 in comparison with other globular clusters and theoretical models.
Several interesting features are seen in Figure 7.
Figure 7(a) illustrates that six SX Phoenicis stars in M53 show a very good correlation
between the period and mean $V$ magnitude,
while the other two stars (SXP2 and SXP8) are in the higher mode than the others.
SXP1, the brightest in the sample, has a period much longer than the rest of the SX Phoenicis stars.
This star is considered to be a member of M53 for three reasons:
first, it is seen inside the blue straggler region in the
color-magnitude diagram; secondly, it is located closer to the center of M53 in the sky
than SXP4 which is clearly a member of M53;
and thirdly, it has period and $V$ magnitude consistent with those of other SX Phoenicis stars
in M53.
We derive a period - luminosity relation using these six stars with $-1.36 < Log P[d]< -1.15$ :
\begin{equation}
<V>=-3.010(\pm0.262)Log P + 15.310(\pm0.048 ) 
\end{equation}
with an rms scatter of 0.038.
This result corresponds to $<M_V>=-3.010 Log P - 1.070$ for an adopted distance modulus of
$(m-M)_V=16.38$ (Harris 1996).

In Figure 7(a) we compare the SX Phoenicis stars in M53 and NGC 5466.
The results for NGC 5466
were also obtained using the same procedures as for M53
from the similar data so that they can
be compared directly with those of M53 \citep{jeo03}.
We adopted distance modulus of $(m-V)_0=16.15$ and zero foreground reddening for NGC 5466
\citep{har96}.
It is shown clearly that the P-L relation of the SX Phoenicis stars in NGC 5466 agrees
very well with that of M53.
The P-L relation for NGC5466 has a slope very similar to that for M53
and a zero point slightly brighter  than that for M53.
Most of the SX Phoenicis stars in NGC 5466 are considered to
be at the fundamental pulsating mode (see \citet{jeo03} for details).
Therefore it is concluded that all SX Phoenicis stars (except for SXP2 and SXP8) in M53
are probably at the fundamental pulsating mode.
Then SXP2 and SXP8 are probably at the first
overtone mode and at the second overtone mode, respectively.
Interestingly the amplitudes of SX Phoenicis stars in NGC 5466 are much larger than those
of M53 stars. This will be discussed in detail in \citet{jeo03}.

In Figure 7(b) we compare the SX Phoenicis stars in M53 and M55 where
as many as 24 SX Phoenicis stars were discovered recently by \citet{pyc01}.
We adopted  distance modulus of $(m-V)_V=13.86$ \citep{pyc01} 
and foreground reddening of $E(B-V)=0.07$ for M55 \citep{har96}.
Note that \citet{pyc01} derived, using the
HIPPARCOS data for SX Phoenicis star itself, a distance modulus,
$(m-V)_V=13.86 \pm 0.25 $, which is very similar to the value given by
\citet{har96}, $(m-V)_V=13.87$.
\citet{pyc01} used SX Phoenicis stars in M55 to estimate the slope of the P-L relation
and used HIPPARCOS data for SX Phoenicis star itself to calibrate the zero point,
deriving a P-L relation for the fundamental mode: $<M_V>=-2.88 Log P - 0.77$.
Figure 7(b) shows that the P-L relation for M53 agrees approximately with that for the
fundamental mode of SX Phoenicis stars in M55
and that the scatter of the P-L relation for M53 is much smaller than that for M55.

In Figure 7(c) we compare the SX Phoenicis stars in M53 and $\omega$ Centauri where
as many as 34 SX Phoenicis stars were discovered by \citet{kal96, kal97}.
Using the data for $\omega$ Centauri, \citet{mcn00} derived steep P-L relations for the fundamental
mode and first overtone mode,
$<V>=-4.66 Log P +11.21$ and $<V>=-4.26 Log P +11.38$, respectively.
We adopted  distance modulus of $(m-V)_V=14.05\pm 0.11$
and foreground reddening of $E(B-V)=0.13$ for $\omega$ Centauri given recently by \citet{tho01} who
used a detached eclipsing binary in $\omega$ Centauri.
These values are very similar to those in \citet{har96}, $(m-V)_V=13.97$
and $E(B-V)=0.12$.

Figure 7(c) shows that the P-L relation for M53 agrees approximately with that for the
SX Phoenicis stars in $\omega$ Centauri which were considered to be
at the first overtone mode by \citet{mcn00} (called A-sequence hereafter).
Note that the steep slope for the first overtone mode in $\omega$ Centauri is based on the data
for the period range of $\log P <-1.3$, where five stars at the fundamental mode in M53
show a similar behavior.
The SX Phoenicis stars at the fundamental mode in $\omega$ Centauri (called B-sequence hereafter)
are 0.4$\sim$0.5 mag fainter than those in M53.

This result is very intriguing and the reasons for this discrepancy are not clear.
$\omega$ Centauri has been known to be a very unique globular cluster
among the Galactic globular clusters in several aspects:
it is the most massive globular cluster in the Milky Way;
it has multiple stellar populations
with a wide spread in metallicity from [Fe/H]=--2.0 to -- 0.6 dex
(the metallicity of the main population is [Fe/H] $\sim -1.6$ dex) ;
and it has a very elongated structure.
These features lead to a suggestion that it may be a remnant core of dwarf spheroidal galaxy
which formed via merging of several components (\citet{gne02, fer02} and references therein).
This peculiarity of $\omega$ Centauri may be related with the fact that
it is the only cluster where more first overtone mode SX Phoenicis stars 
were found than the fundamental mode stars (or in comparable number).
Note that much fewer first overtone mode stars were found than the fundamental mode stars
in globular clusters.

We checked several possibilities to understand this puzzling result:
(1) If the A-sequence and the B-sequence are, respectively, indeed first overtone mode and
the fundamental mode as considered by \citet{mcn00}
(and if the P-L relation of SX Phoenicis stars is universal),
then the distance modulus of either $\omega$ Centauri  or M53 must be wrong by
as much as 0.4$\sim$0.5 mag. This is very unlikely, considering the
good agreement among the different estimates of the distance to these objects;
(2) The A-sequence may be fundamental mode rather than first overtone mode. Then it is not possible
to explain the existence of the B-sequence
without assuming that the B-sequence is another fundamental mode
behind the A-sequence. This is also unlikely for the same reason as (1);
(3) The A-sequence and the B-sequence may not be two independent modes.
Instead both may belong to the same fundamental mode with a large dispersion.
This is contrary to the result given by \citet{mcn00}.
However, it happens to be consistent with
the theoretical P-L relation for the fundamental mode given by \citet{tem02},
as shown in Figure 7(d).
Therefore this problem remains to be explained by further studies.

In Figure 7(d) we compare the P-L relation for M53 with empirical relations given by
\citet{mcn97, mcn02} and with theoretical relations given by \citet{san01}
and \citet{tem02}.
The slope of the P-L relation for the fundamental mode of M53 is very similar
to those in theoretical ones (--3.05 \citep{san01} and --3.14 \citep{tem02} ).
The six SX Phoenicis stars in M53 are located at the right side of the blue edge of
the fundamental mode relation given by \citet{san01}, consistent with the
conclusion that these stars are indeed the fundamental modes.
On the other hand, the zero point of the M53 relation is about 0.4 mag brighter
than that of the theoretical relation given by \citet{tem02}.
The reason for this discrepancy is not known at this moment.
The empirical P-L relations given by \citet{mcn97, mcn02}
($<M_V>= -3.725  \log P - 1.933$ $<M_V> = -4.14 \log P - 2.46$ ) are
steeper than those for M53 and the theoretical relations given 
by \citet{san01} and \citet{tem02}.



%
\placefigure{fig7}

\section{Summary}

We present time-series $BV$ CCD photometry of the metal-poor globular cluster M53.
we have discovered eight  new SX Phoenicis type stars (labeled from SXP1 to SXP8).
Physical parameters of these stars are summarized in Tables 2 and 3.
All the new SX Phoenicis stars are located in the blue straggler star
region of a color-magnitude diagram of M53.
One of these stars (SXP2) is found to have very closely
separated pulsation frequencies: $f_1/f_2 = 0.9595$
where $f_1$ and $f_2$ are primary and secondary frequencies.
This may be due to excitation of non-radial modes.

Considering the position in the period-$V$ magnitude diagram ,
six of these  SX Phoenicis stars are identified to be pulsating
at the fundamental mode,
and SXP2 and SXP3 are probably at the first overtone and second overtone modes.
We derive a period - luminosity relation for the fundamental mode
for the period range of $-1.36 < \log P[d]< -1.15$ :
$<V>=-3.010(\pm0.262)Log P + 15.310(\pm0.048 )$ with an rms scatter of 0.038,
corresponding to $<M_V>=-3.010 Log P - 1.070$ for an adopted distance modulus
of $(m-M)_V=16.38$ (Harris 1996).

\acknowledgments
Authors are very much grateful to Dr. D. H. McNamara, the referee, who provided
very useful comments to improve this paper. M.G.L thanks the director and staff of
the Department of Terrestrial Magnetism, the Carnegie Institution of Washington,
for their kind hospitality.
M.G.L. was supported in part by the Korean Research Foundation Grant
(KRF-2000-DP0450).

\clearpage

\plotone{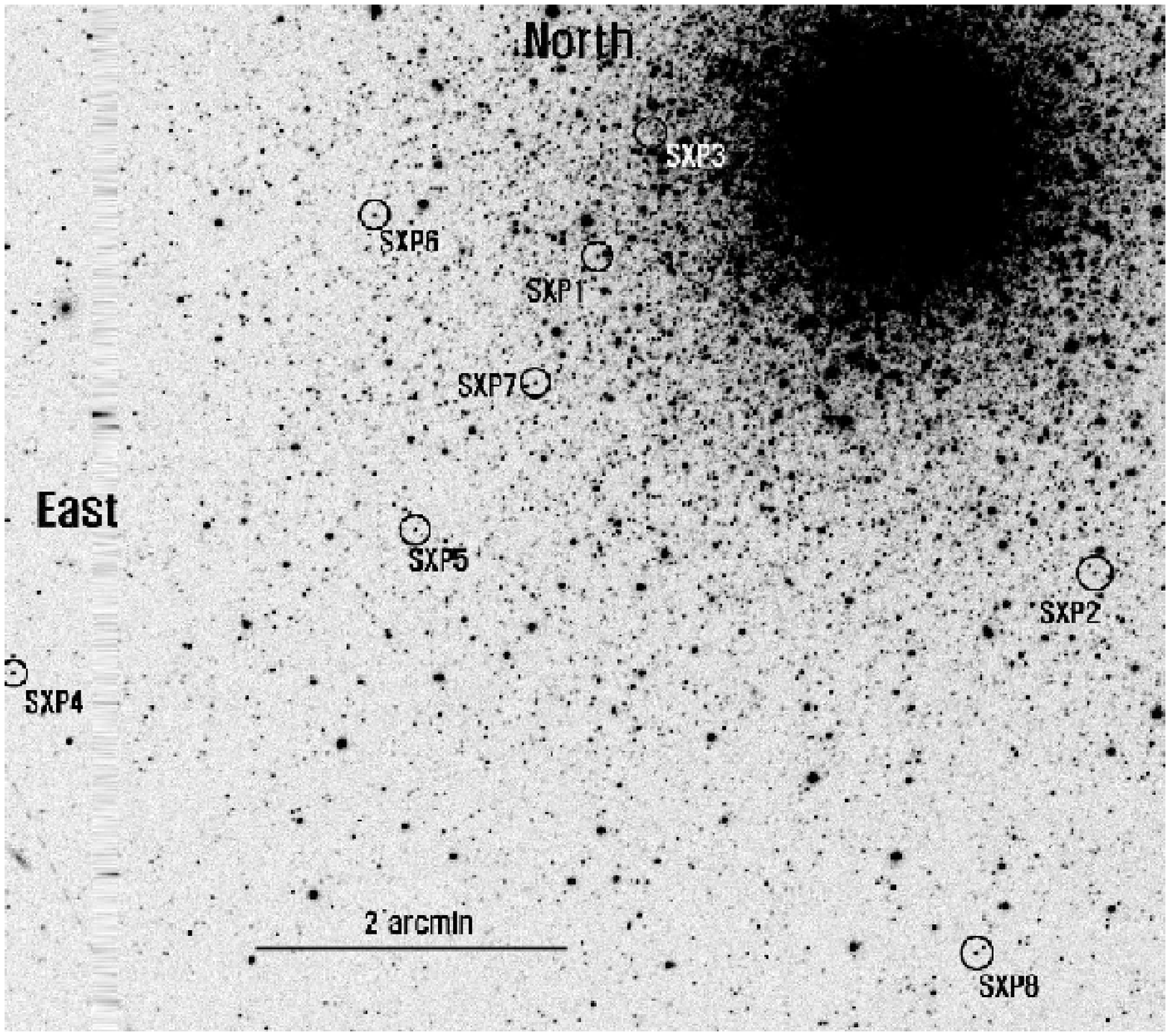}
\figcaption[Jeon.fig01.ps]{ 
A greyscale map of a $V$-band CCD image of the globular cluster M53.
This image shows a $7\farcm6\times6\farcm7$ field in
 the south-east region  of the cluster, out of the observed field of
$11\farcm6\times11\farcm6$.
Eight  new SX Phoenicis stars are labeled from SXP1 to SXP8.
\label{fig1}}

\plotone{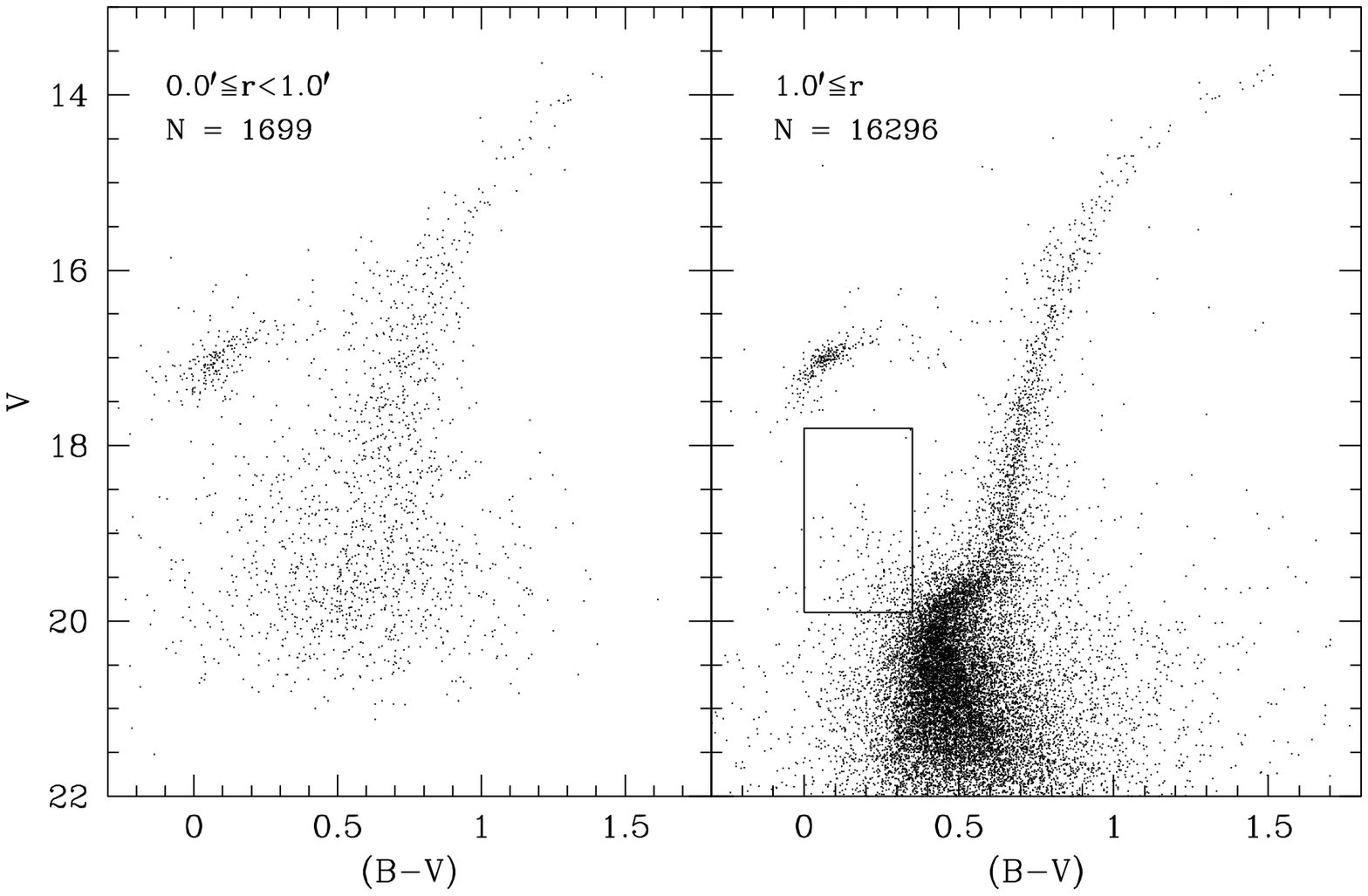}
\figcaption[Jeon.fig02.ps]{
Color-magnitude diagrams of M53.
The left panel is for a central region at r$ < $1$\farcm0$,
and the right panel is for an outer region at r$ \ge$1$\farcm$0.
The box represents a blue straggler region.
\label{fig2}}

\plotone{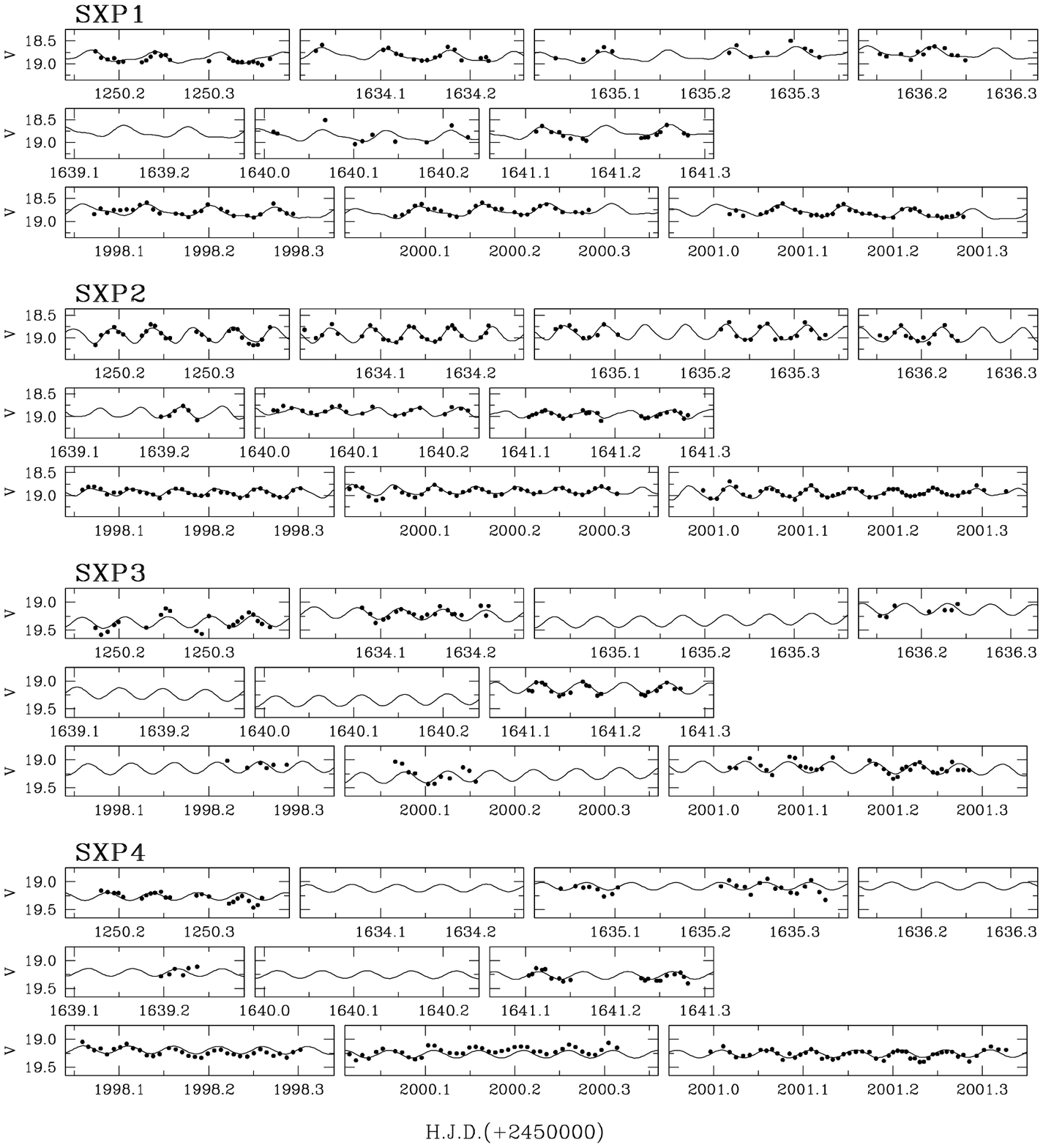}
\clearpage
\plotone{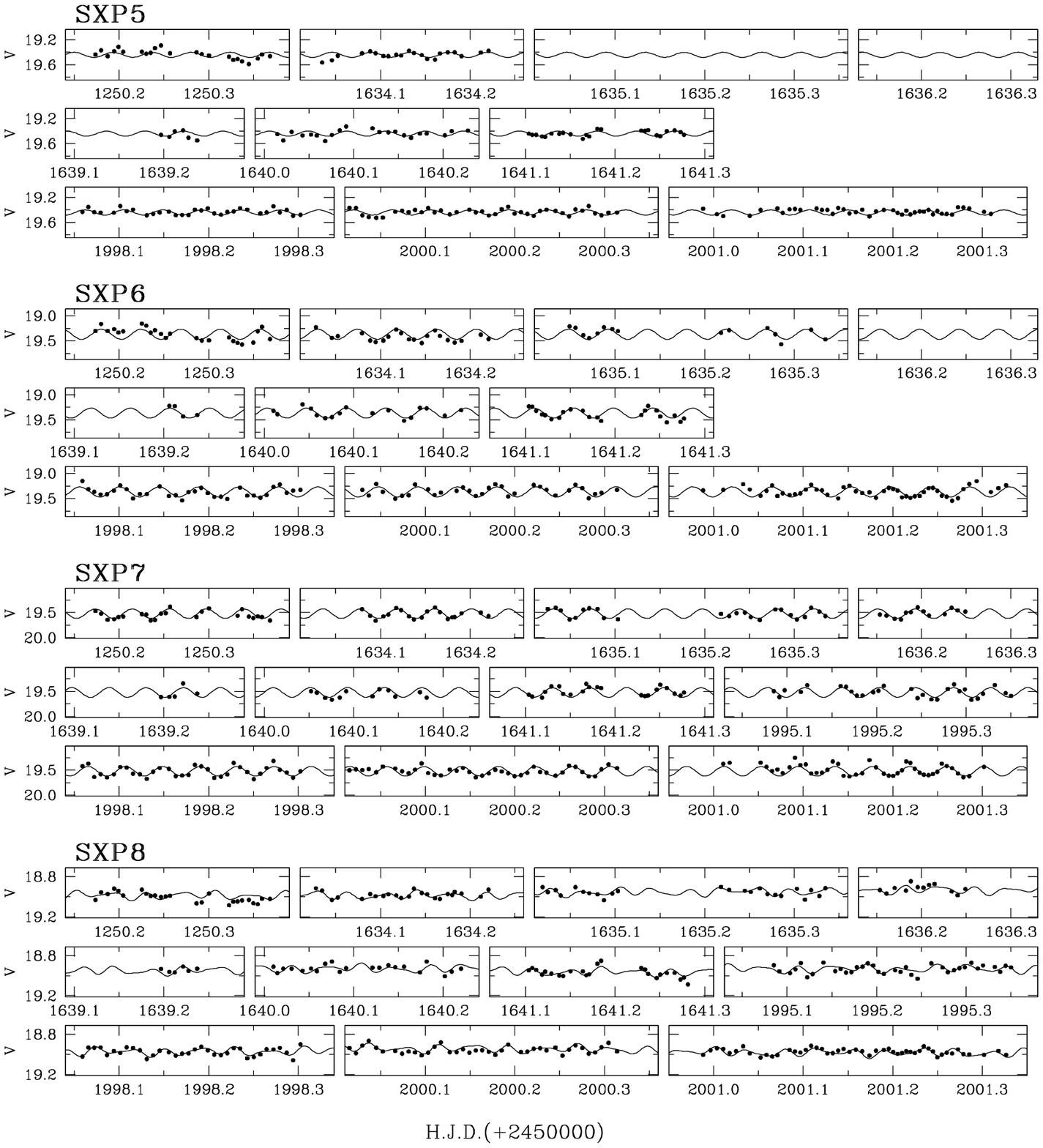}
\figcaption[Jeon.fig03a.ps]{ Observed light curves
(dots) for eight new SX Phoenicis stars.
Synthetic light curves (solid lines) obtained from the
multiple-frequency analysis (see Table 3) are superimposed on the
data. \label{fig3}}

\clearpage

\plotone{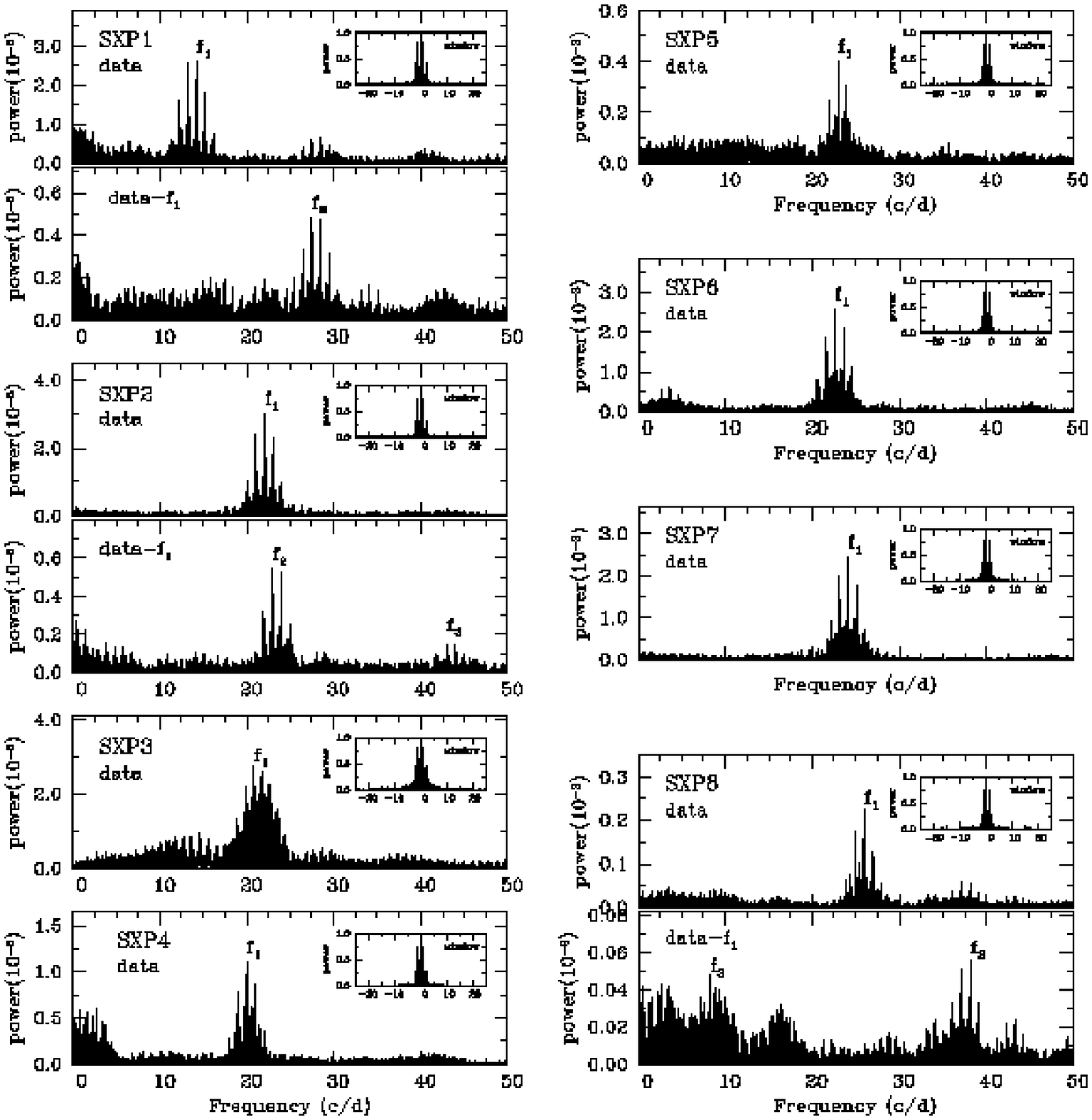}
\figcaption[Jeon.fig04.ps]{ Power spectra of eight new SX Phoenicis stars.
Window spectra are shown in the small box within each panel.
\label{fig4}}

\plotone{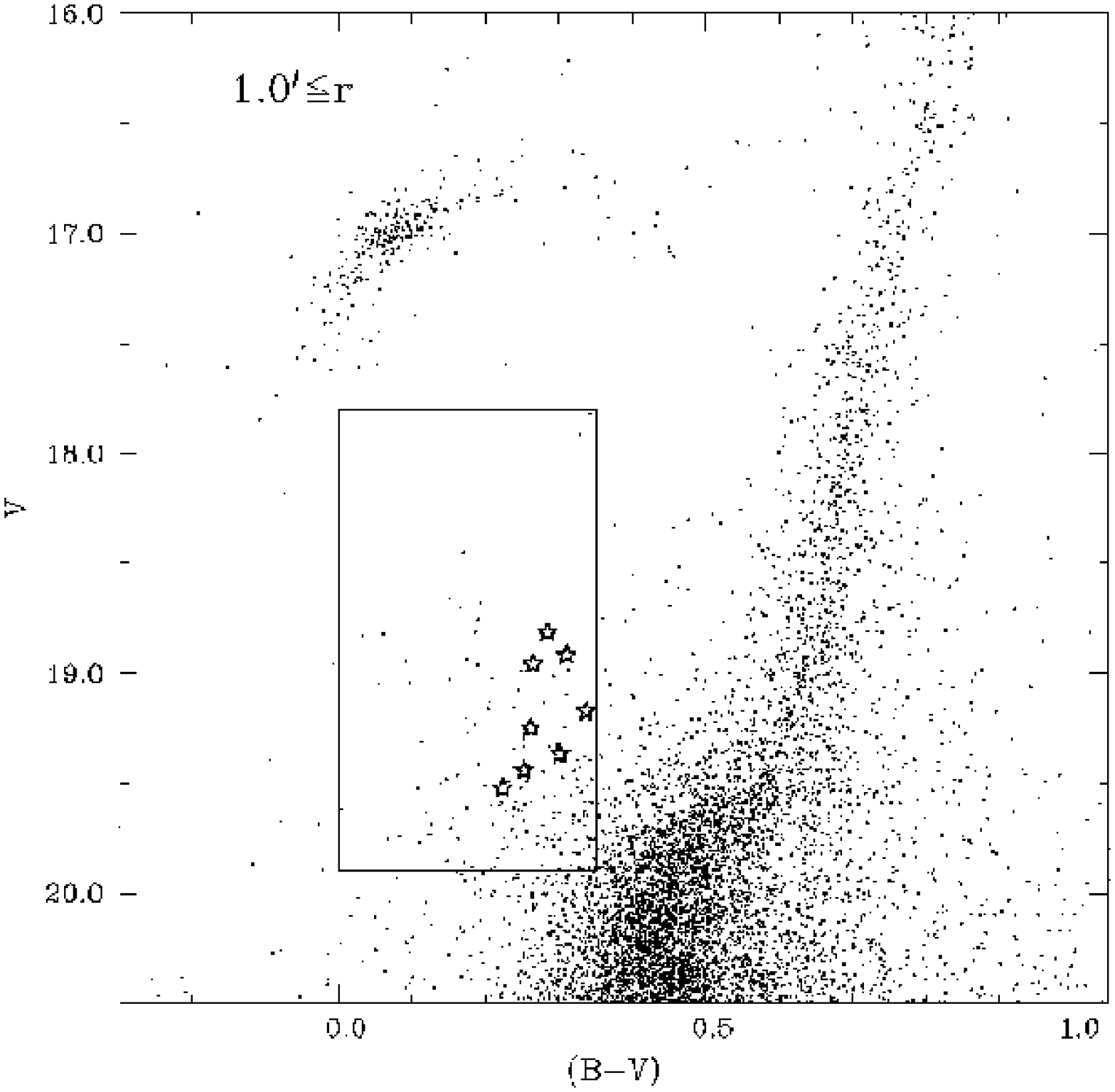}
\figcaption[Jeon.fig05.ps]{ Positions of  eight new SX Phoenicis stars
in the color-magnitude diagram of M53. Note that all they are located in
the blue straggler region.
\label{fig5}}

\plotone{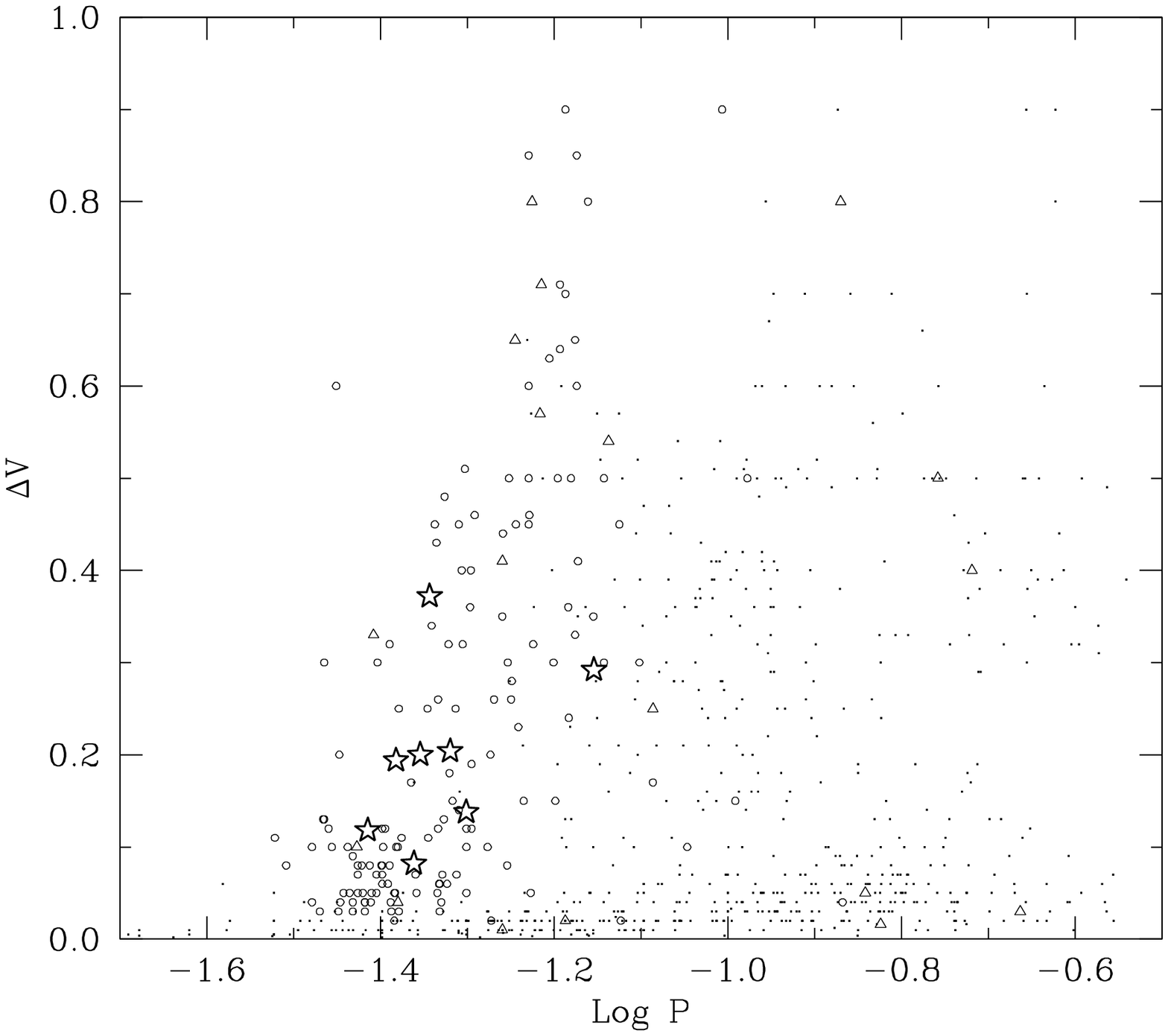}
\figcaption[Jeon.fig06.ps]{ $V$ amplitude versus  period diagram.
Star symbols denote eight new SX Phoenicis stars in M53,
triangles represent field SX Phoenicis stars,
open circles indicate SX Phoenicis stars in other globular clusters,
and dots denote $\delta$ Scuti stars.
\label{fig6}}

\plotone{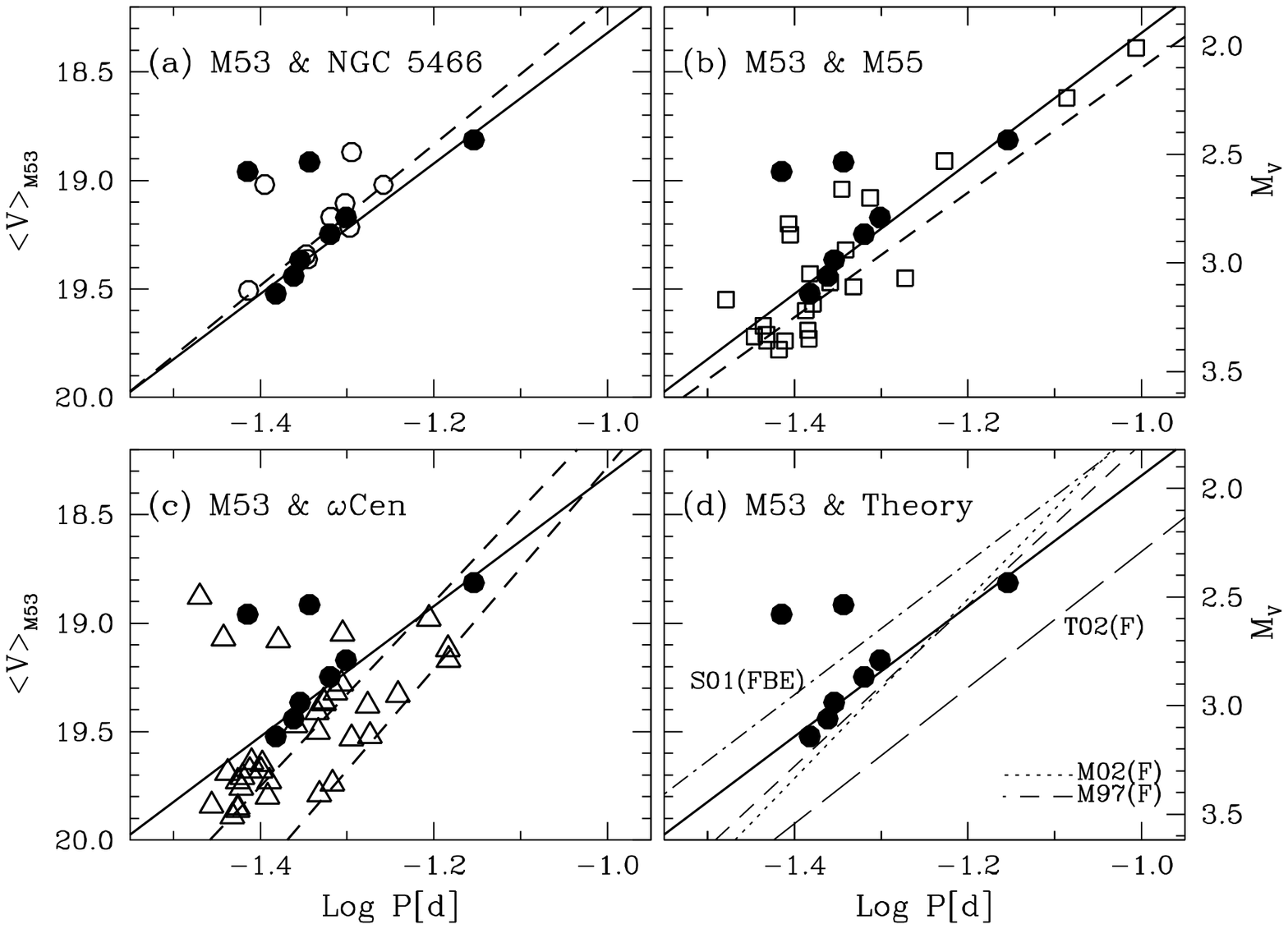}
\figcaption[Jeon.fig07.ps] { Mean magnitude $<$$V$$>$ versus  period diagram.
Filled circles represent the SX Phoenicis stars in M53. The solid line
represent a linear fit to the data for the fundamental mode of M53.
(a) Comparison with the SX Phoenicis stars in NGC 5466 (open circles).
The dash line represents a linear fit to the data of the fundamental mode of
NGC 5466 (Jeon et al. 2003).
(b) Comparison with the SX Phoenicis stars in M55 (open squares).
The dash line represents a linear fit to the data for the fundamental mode of M55
(Pych et al 2001).
(c) Comparison with the SX Phoenicis stars in $\omega$ Centauri (open triangles).
The two dashed lines represents linear fits to the data for the A-sequence (left)
and the B-sequence (right) of  $\omega$ Centauri.
(d) Comparison with empirical P-L relations (short dashed lines) given 
by McNamara (1997, 2002)
and with theoretical P-L relations. The long dashed line represents a
theoretical relation for the fundamental mode given by Templeton et al. (2002).
The dot-dashed lines, respectively, represent the blue edge of the fundamental mode,
the first overtone mode and the second overtone mode (from the bottom to top)
given by Santolamazza et al. (2001).
\label{fig7}}
\clearpage

\begin{deluxetable}{lrcrrrc}

\tablecaption{Observation Log. \label{tbl-1}} \tablewidth{0pt}
\tablecolumns{7}
\tablehead{
\colhead{Date} & \colhead{Start HJD} & \colhead{Duration} & \colhead{ $N_{obs}$  } &
\colhead{Seeing} & \colhead{Exposure Time} & \colhead{Remarks} \\
\colhead{(UT)} & \colhead{(2,450,000+)} & \colhead{(hours)} & \colhead{  } &
\colhead{(arcsec)} & \colhead{(seconds)} & \colhead{}
} \startdata

1999~~3~12 & 1250.173(V) & 4.7 & 30(V) & 1.2$\sim$2.4(V) & 300(V)  \\
2000~~3~30 & 1634.007(V) & 5.1 & 38(V) & 1.2$\sim$2.3(V) & 180, 240(V) & standard stars \\
             & 1634.022(B) &  & 12(B) & 1.3$\sim$2.0(B) & 300, 450(B)   \\
2000~~3~31 & 1635.018(V) & 7.6 & 46(V) & 2.1$\sim$3.4(V) & 210, 180(V)  \\
             & 1635.029(B) &  & 14(B) & 2.3$\sim$3.1(B) & 300(B)            \\
2000~~4~~1 & 1636.153(V) & 2.3 & 16(V) & 2.2$\sim$2.4(V) & 420, 210(V)  & thin cloud\\
            & 1636.165(B) &  & 1(B)  & 2.6(B)          & 300(B)        \\
2000~~4~~4 & 1639.196(V) & 1.0 & 6(V)  & 2.2$\sim$2.4(V) & 210(V)     \\
            & 1639.201(B) &  & 3(B)  & 2.3$\sim$3.0(B) & 300(B)       \\
2000~~4~~5 & 1640.010(V) & 5.2 & 38(V) & 2.8$\sim$3.6(V) & 210, 300(V)  \\
            & 1640.025(B) &  & 11(B) & 3.0$\sim$3.7(B) & 300(B)     \\
2000~~4~~6 & 1641.103(V) & 4.3 & 34(V) & 1.6$\sim$2.1(V) & 210, 180(V) & thin cloud\\
            & 1641.133(B) &  & 4(B)  & 1.8$\sim$1.9(B) & 300(B)      \\
2001~~3~26 & 1995.085(V) & 6.4 & 42(V) & 3.1$\sim$4.0(V) & 400(V)     & bad seeing\\
2001~~3~29 & 1998.058(V) & 5.9 & 34(V) & 1.9$\sim$2.8(V) & 350 $\sim$ 500(V)  \\
2001~~3~31 & 2000.000(V) & 7.5 & 42(V) & 1.8$\sim$2.7(V) & 500(V)      \\
2001~~4~~1 & 2000.988(V) & 8.1 & 53(V) & 1.2$\sim$2.3(V) & 300 $\sim$ 600(V)  \\

\enddata
\end{deluxetable}

\clearpage

\begin{deluxetable}{lcccc}

\tablecaption{Observational Parameters of the Eight SX Phoenicis Stars. \label{tbl-2}} \tablewidth{0pt}
\tablecolumns{5}
\tablehead{
\colhead{Name} & \colhead{RA(J2000.0)} & \colhead{Dec(J2000.0)} & \colhead{$<$$V$$>$ } &
\colhead{$<$$B$$>$$-$$<$$V$$>$}
} \startdata
SXP1 &  13 13 03.38 & 18 09 25.3 & 18.814  &   0.284 \\
SXP2 &  13 12 49.71 & 18 07 26.0 & 18.915  &   0.310 \\
SXP3 &  13 13 01.95 & 18 10 13.4 & 19.248  &   0.261 \\
SXP4 &  13 13 19.09 & 18 06 40.7 & 19.171  &   0.337 \\
SXP5 &  13 13 08.21 & 18 07 38.7 & 19.441  &   0.252 \\
SXP6 &  13 13 09.44 & 18 09 40.1 & 19.366  &   0.301 \\
SXP7 &  13 13 05.01 & 18 08 36.1 & 19.522  &   0.223 \\
SXP8 &  13 12 52.77 & 18 04 58.4 & 18.959  &   0.264 \\

\enddata
\end{deluxetable}

\clearpage

\begin{deluxetable}{ccccrrrcc}
\tablecaption{Pulsating Properties of the Eight SX Phoenicis Stars.
\label{tbl-3}} \tablewidth{0pt} \tablecolumns{9} \tablehead{
\colhead{Name} & \colhead{Value} &
\colhead{Frequency\tablenotemark{{a,b}}} & \colhead{Amp.\tablenotemark{b}}   &
\colhead{Phase\tablenotemark{b}}  &
\colhead{S/N\tablenotemark{c}} &  \colhead{Mode} & \colhead{Remarks}
} \startdata
SXP1 & $f_1$  & 14.2600 & 0\fm103  &   0.6873&  9.0 & F         \\
     & $f_2$  & 27.5256 & 0\fm043  & --1.1868&  5.1 & 2$f_1$    & $f_2-2f_1=-0.9944$\\
\cline{1-8}
SXP2 & $f_1$  & 22.0450 & 0\fm118  & --0.0736& 14.4 & 1H        & the highest amp. \\
     & $f_2$  & 22.9750 & 0\fm047  &   1.5520&  7.2 & Non-radial& $f_1/f_2=0.9595$ \\
     & $f_3$  & 43.0190 & 0\fm021  & --1.3179&  3.8 & 2$f_1$    & $f_3-2f_1=-1.0630$\\
\cline{1-8}
SXP3 & $f_1$  & 20.8796 & 0\fm102  &   4.2981&  7.0 &  F\\
\cline{1-8}
SXP4 & $f_1$  & 20.0114 & 0\fm069  &   3.1873&  7.9 &  F\\
\cline{1-8}
SXP5 & $f_1$  & 22.9880 & 0\fm041  &   0.3222&  6.0 &  F\\
\cline{1-8}
SXP6 & $f_1$  & 22.6000 & 0\fm100  &   2.6747& 12.1 &  F\\
\cline{1-8}
SXP7 & $f_1$  & 24.1024 & 0\fm097  &   2.0239& 14.4 &  F\\
\cline{1-8}
SXP8 & $f_1$  & 25.9800 & 0\fm030  &   3.1414&  8.2 & 2H  \\
     & $f_2$  & 38.1912 & 0\fm015  & --1.5238&  4.4 & Non-radial?     \\
     & $f_3$  &  8.2322 & 0\fm014  & --0.7535&  4.3 & Non-radial?     \\

\enddata

\tablenotetext{a}{In cycles per day. }
\tablenotetext{b}{$V = Const + \Sigma_j A_j \cos \{2 \pi
          f_j (t - t_0) + \phi_j\},~~ t_0 =$ HJD 2,450,000.00. }
\tablenotetext{c}{Amplitude signal-to-noise ratio introduced by Breger et al. (1993). }

\end{deluxetable}

\clearpage

\end{document}